\documentstyle[epsfig,floats,prb,aps]{revtex}

\begin{document}

\title{\bf Model calculations of the proximity effect in finite multilayers}

\author{C. Ciuhu and A. Lodder}

\address{Faculty of Sciences / Natuurkunde en Sterrenkunde, Vrije Universiteit,
         De Boelelaan 1081,\\ 1081 HV Amsterdam, The Netherlands}

\date{\today}
\maketitle

\begin{abstract}
\normalsize{

The proximity-effect theory developed by Takahashi and Tachiki \cite{taka}
for infinite multilayers is applied to multilayer systems with a finite
number of layers in the growth direction. The purpose is to investigate
why previous applications to infinite multilayers fail to describe the
measured data satisfactorily \cite{rutg}. 
Surface superconductivity may appear, depending
on the thickness of the covering normal metallic $N$ layers on both
the top and the bottom. The parameters used  are
characteristic for V/Ag and Nb/Pd systems. The nucleation process is
studied as a function of the system parameters.}
\end{abstract}

\section{Introduction}
\label{intro}

An alternating sequence of superconductor and normal metal layers (S/N) 
generates a system whose properties have raised both theoretical and
experimental interests. The dirty-limit theory of Takahashi and
Tachiki\cite{taka} is meant to describe such systems. However, it 
has some insufficiencies in explaining experimental results such as the 
dimensional crossover in $H_{c2,\parallel}(T)$ curves,\cite{rutg,rutger}
displaying the upper critical field applied parallel to the layers
as a function of the temperature. Since up to now exact 
calculations using this theory were done for infinite periodic systems, we 
extend the calculations to finite multilayers. 
A marked difference is found in the proximity effect on the pair
amplitude for the upper critical magnetic fields $H_{c2,\perp}$ and $H_{c2,\parallel}$.
In trying to cover as many aspects as possible which underlie the physics in
such systems, we are hoping to prepare a theoretical
explanation of the experimentally observed dimensional crossover in the
parallel upper critical field $H_{c2,\parallel}$ versus temperature, which
improves upon the at present available explanation, being a rather
artificial {\it ad hoc} explanation.\cite{rutg}
We apply the theory to real systems such as finite multilayers of V/Ag.

Section \ref{proxim} gives a brief survey of the Takahashi-Tachiki theory. 
In Section \ref{theory} we calculate the spatial distribution of the
pair amplitude, showing features of
the proximity effect, which are typical for layered systems.
Two distinct situations are
treated separately, namely one in which the magnetic field applied to the
system is perpendicular to the layers and the other one
when the field is parallel to the layers.
Section \ref{nucleation} is devoted to nucleation properties, which
form an intriguing particularity of
superconductivity when a parallel magnetic field is applied.
The manifestation of surface superconductivity and related 
problems, such as the 
influence and role played by the boundary conditions, are treated.
In Section \ref{parameters} we will present a detailed description of the 
dependence of the proximity effect on the parameters of the system. 
The results obtained restore confidence in the applied
microscopic theory for dirty multilayers.

\section{The Takahashi-Tachiki theory}
\label{proxim}

In inhomogeneous systems such as S/N multilayers, the superconducting
properties are influenced by the proximity of different materials. 
One can think of
non-superconducting materials, or of superconducting materials with a lower
$T_{c}$.
A finite Cooper pair amplitude is induced in the less- or non-superconducting
material. 
This proximity effect leads to a reduction of the critical temperature $T_{c}$
compared to that of a bulk superconductor.

We will apply the description developed by Takahashi and Tachiki,
which has been devised to be valid for inhomogeneous superconductors
in the dirty limit.\cite{taka}
We now give a brief summary of the main equations, by that also
defining the system parameters to be used.

The theory starts from the Gor'kov equation \cite{gork}
for the pair potential $\Delta( {\bf r})$ 
\begin{equation}
\label{gorkov}
\Delta ({\bf r})=V( {\bf r})kT \sum_{ \omega} \int d^{3} {\bf r'}
Q_{ \omega}( {\bf r,r'})\Delta ( {\bf r'}),
\end{equation}
in which the material parameter $V( {\bf r})$ is the space dependent BCS
coupling constant and the integration kernel $Q_{\omega}$ obeys
a Green's function like equation 
\begin{equation}
\label{green}
[2|\omega |+L({\bf \nabla})]Q_{\omega }({\bf r,r'})=2\pi N({\bf r})\delta
({\bf r-r'}).
\end{equation}
The differential operator is defined by
\begin{equation}
\label{operator}
L({\bf \nabla })=- \hbar D({\bf r})({\bf \nabla} -\frac {2ie}{\hbar c }{\bf A}
({\bf r}))^2.
\end{equation}
The material parameters $N({\bf r})$ and $D({\bf r})$ are
the electronic density of states at the Fermi energy
and the diffusion coefficient, respectively. $V({\bf r})$,
$N({\bf r})$ and $D({\bf r})$
are constant in each single layer. At the interfaces
de Gennes boundary conditions are imposed,\cite{degen}
which require the continuity of
$\frac{F({\bf r})}{N({\bf r})}$ and $D({\bf r})({\bf \nabla} -\frac {2ie}
{\hbar c }{\bf A}({\bf r}))F({\bf r})$, where the pair amplitude
$F({\bf r})$ is related to the gap function $\Delta ({\bf r})$ through
\begin{equation}
\label{pair-gap}
\Delta ({\bf r})=V({\bf r})F({\bf r}).
\end{equation}

Takahashi and Tachiki provide a way of solving Eqs.
(\ref{gorkov}) and (\ref{green}) 
by developing the kernel $Q_{\omega }({\bf r,r'})$ and the pair function
$F({\bf r})$ in terms of a complete set of eigenfunctions of the operator
$L({\bf \nabla })$. These eigenfunctions are labeled by the parameter
$\lambda$, and the eigenvalues are $\epsilon_{\lambda }$. They are solution 
of the eigenvalue problem
\begin{equation}
\label{lambda}
L(\nabla)\Psi_{\lambda }= \epsilon_{\lambda }\Psi_{\lambda }.
\end{equation}
The requirement of the existence of a solution for Eq. (\ref{gorkov})
leads to the equation
\begin{equation}
\label{det}
\det|\delta _{\lambda \lambda '}-2\pi kT\sum _{\omega }\frac{1}{2|\omega |+
\epsilon_{\lambda }}V_{\lambda \lambda '}|=0.
\end{equation}
For finite multilayers covered by infinite N layers, 
the boundary conditions at infinity, imposed in determining the eigenfunctions
$\Psi _{\lambda}(x,y,z)$ of the differential operator $L({\bf \nabla })$, are
such that $\Psi _{\lambda}(z \rightarrow \pm\infty)=0$, which
is in line with $F(z \rightarrow \pm\infty)=0$.
As usual for these type of layered systems, the growth direction
coincides with the $z$ direction.
For finite multilayers in vacuum, we apply de Gennes boundary conditions, 
since they insure that there is no current flow through the interface 
between the multilayer and the vacuum. These boundary conditions read 
$\frac{\partial \Psi _{\lambda}(x,y,z)}{\partial z}=0$ at the interface
with the vacuum. 
In the absence of a magnetic field, the solution of Eq. (\ref{det}) giving the
largest value for the critical temperature is the physical one.
In the presence of a field, solving this equation allows us to derive
the $H_{c2} (T)$ curves.
The temperature at which $H_{c2}\rightarrow 0$ is $T_c$.

\section{The perpendicular and parallel upper critical magnetic fields}
\label{theory}

\begin{figure}[htb]
\centerline{\epsfig{figure=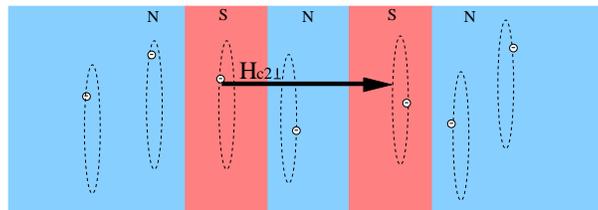,angle=-90,width=8.0cm}}
\caption[]{The movement of the electrons in the system, when a perpendicular
magnetic field is applied.}
\label{perpfig}
\end{figure}

We consider a multilayer system such as the one depicted in
FIG. \ref{perpfig}.
A certain
number of finite-thickness S/N layers are covered by infinitely thick external
$N$-layers in the growth direction $z$. We will
vary the number of internal layers, and show results for systems with
1, 5, 9, and 69 internal layers respectively.
A magnetic field of magnitude 1 kGauss is applied 
in the growth direction, so 
perpendicular to the layers. 
The computation and analysis of the pair function will lead us to some
interesting conclusions related to the microscopic aspects of the proximity
effect.

For a perpendicular orientation of the field we can choose ${\bf A}({\bf r})=
(0,Hx,0)$. By that,
applying the method of separation of variables, we can write
the solution $\Psi_{\lambda}$ of Eq. (\ref{lambda}) as
\begin{equation}
\label{psi-lambda}
\Psi_{\lambda}=w_{a_{x}}(x)e^{ik_{y}y}\Phi_{\lambda}(z)
\end{equation}
The solutions in the $x$ direction are parabolic cylinder functions,
which for the present case are equivalent to one-dimensional
harmonic oscillator functions. 
The system is infinite and uniform in the $x$ and $y$ directions, so that the 
ground state has $a_x=-\frac{1}{2}$ and arbitrary $k_y$. With
$\Psi_{\lambda}$ given by Eq. (\ref{psi-lambda}), Eq. (\ref{lambda}) reduces to
\begin{equation}
\label{perp-field}
-\frac{d^2}{dz^2}\Phi _{\lambda}(z)=k_{\lambda}^{2}(z)\Phi _{\lambda}(z),
\end{equation}
with
\begin{equation}
\label{perp-f}
k_{\lambda}^{2}(z)=\frac{\epsilon_{\lambda}}{\hbar D(z)}-\frac{1}{\xi^2},
\end{equation}
in which the magnetic coherence length $\xi  =\sqrt{\frac{\hbar c}{2eH}}$. 

\begin{figure}[htb]
\centerline{\epsfig{figure=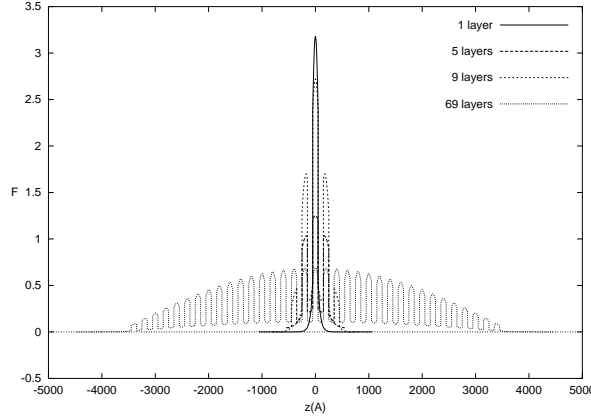,angle=-90,width=8.0cm}}
\caption[]{The pair function in perpendicular magnetic field as a function of
the coordinate $z$ in the growth direction. The thickness
of both N and S layers are $d_S=d_N=100$\AA. The
magnetic field applied is $H_{\perp}$=1 kGauss.}
\label{pairperp}
\end{figure}

In FIG. \ref{pairperp} the pair function is shown for four systems
with a different number of internal layers.
The system parameters used are specific for V/Ag multilayers 
\cite{kanoda1,kanoda2,kanoda3,kanoda4}.
Two features of the proximity effect come out clearly.  One is that the
proximity of the S and N layers leads to the oscillating behavior, 
in which the minima correspond to the N layers, while the maxima occur within 
the S layers. The other
one is an overall proximity effect due to the infinite exterior $N$-layers.
The 69-layers result
can be considered as to stand for the limit of a large number
of layers. The influence of the outer layers is not noticed any more
in the middle of the system, the behavior here becoming similar to that of an
infinite periodic system, described by Koperdraad et al.
\cite{rutg}

\begin{figure}[htb]
\centerline{\epsfig{figure=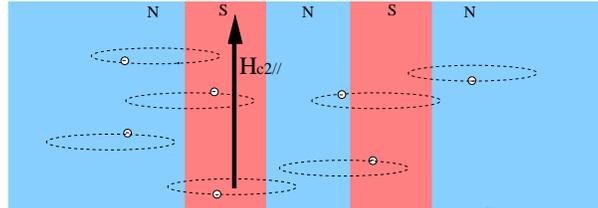,angle=-90,width=8.0cm}}
\caption[]{The movement of the electrons in the system, when a parallel
magnetic field is applied.}
\label{para}
\end{figure}

Turning to the parallel field case, we first show in FIG. \ref{para}
in an intuitive way how the electrons move. This leads to a movement of the
Cooper pairs in which they would extend naturally in the normal layers.
But as they are forbidden there, the concentration of Cooper pairs in
the superconducting layers will be pushed up. As in the perpendicular field
case, the outer $N$ layers extend to infinity.

To compute the pair functions, we go back to Eq. (\ref{lambda}). A parallel
magnetic field lies in the $xy$-plane, and we can choose the vector
potential ${\bf A}({\bf r})$ as $(Hz, 0, 0)$.
Separation of the variables leads to the form
\begin{equation}
\label{psilampar}
\Psi_{\lambda}=e^{ik_{x}x}e^{ik_{y}y}w_{\lambda}(z)
\end{equation}
for $\Psi_{\lambda}$, and the equation for $w_{\lambda}$ can be written as
\begin{equation}
\label{para-f}
(-\frac{d^2}{dz^2}+\frac{(z-z_0)^2}{\xi^4}+k_y^2)w_{\lambda}(z)=\frac{
\epsilon_{\lambda}}{\hbar D(z)}w_{\lambda}(z).
\end{equation}
The point $z_0\equiv \xi^2k_x$ is called the nucleation point. For the
results to be presented in this section the nucleation point is
chosen in the middle of the system, so $z_0 = 0$. In the following
section it will become clear, that this symmetry-inspired choice
is not the only possibility, particularly if the covering $N$ layers
are no longer infinitely thick and the system becomes
really finite in the growth direction.

\begin{figure}[htb]
\centerline{\epsfig{figure=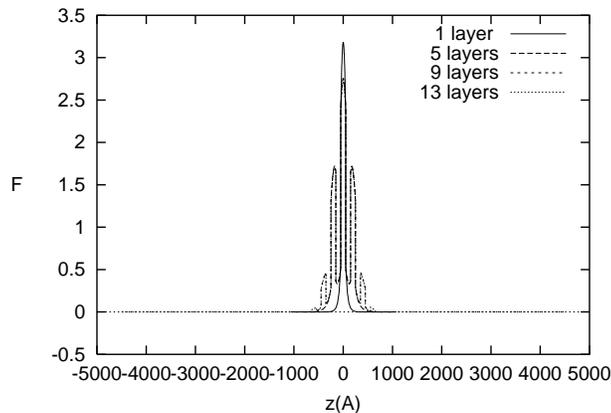,angle=-90,width=8.0cm}}
\caption[]{The pair function for systems with a different number of layers in
a parallel magnetic field, for the layer thickness
$d_S=d_N=100$\AA. The applied magnetic field is
$H_{\parallel}=7$ kGauss.}
\label{pairpara}
\end{figure}

\begin{figure}[htb]
\centerline{\epsfig{figure=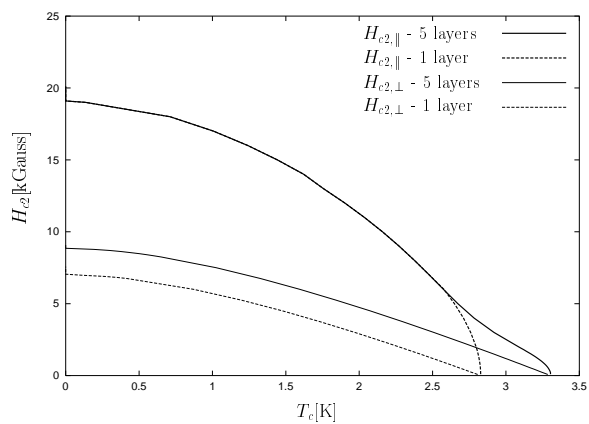,angle=+90,width=8.0cm}}
\caption[]{$H_{c2,\parallel}(T)$ and $H_{c2,\perp}(T)$ curves.}
\label{curves}
\end{figure}

In FIG. \ref{pairpara} we show the pair functions of the systems 
containing 1, 5, 9, and 13 internal layers. The markedly different
behaviour compared to what was found for the perpendicular
field case is certainly due to the anisotropy of the system.
For each system considered, the pair function drops
to zero, except for a few layers, 
centered around the nucleation point $z_0$.
Apparently now the proximity effect, here being equivalent to
the spreading of the pair amplitude, is much weaker than for $H_{c2,\perp}$.
The result shown is
obtained for a magnetic field of $7$ kGauss, which is somewhat larger
than the field used in the perpendicular field case. In the latter case the
features of the pair distribution are insensitive to the
magnitude of the magnetic field. But it appears
that in the parallel field case, the higher the field is,
the more the pair function is concentrated at the center. So one
expects not much difference in the $H_{c2,\parallel}(T)$ curves for
lower temperatures. This is ilustrated nicely by FIG. \ref{curves},
in which these curves are shown for the monolayer and 5-layers systems.
The $H_{c2,\perp}(T)$ curves are shown as well.
The $H_{c2,\parallel}(T)$ curves indeed coincide for higher parallel 
magnetic fields, irrespective to the number of internal layers considered, 
but they are clearly different for a small parallel 
magnetic field. This is related to the fact, that
for small magnetic fields, the pair function is spread over the multilayer,
in a similar way as it is for the perpendicular magnetic field.

The fact that the parallel upper critical field is larger than
the perpendicular field is a consequence of the
difference in the distribution of the pair amplitude for the two field
directions. In the parallel field case, the pairs are
concentrated in the S layers. That is why for this direction a larger
magnetic field is needed
to destroy the superconductivity. A similar argument explains
the fact, that $H_{c2,\perp}$ for five layers is larger than for
one layer. Since for $H_{c2,\perp}$  the pair amplitude is spread over
the layers, a 5-layers system is a better superconductor than
a one-layer system for all temperatures.
Consequently, a one-layer system is expected to have a lower
$T_c$ than a 5-layers system. This is shown clearly by the $H_{c2,\perp}(T)$
curves, and that is why the parallel curves converge to the
perpendicular ones in the zero-field limit.

Now we turn to really finite systems, for which the covering
$N$ layers are finite as well.

\section{Search for the nucleation point}
\label{nucleation}

Measurements of the parallel upper critical magnetic field $H_{c2,\parallel}$ for 
multilayer systems such as V/Ag \cite{kanoda1,kanoda2,kanoda3,kanoda4} show 
interesting features, such as the dimensional crossover in 
$H_{c2,\parallel}(T)$ curves. A generic 
$H_{c2,\parallel}(T)$ curve is divided in three regions. Near $T_c$, there is an 
average 3D behavior, corresponding to a linear dependence of the critical 
magnetic field on the temperature. 
At a lower temperature, at which the magnetic coherence length
becomes of the order of the layer thickness, this behavior is 
replaced with a square-root-like dependence of $H_{c2,\parallel}$ on $T$, 
corresponding to a 2D domain. 
A second dimensional crossover is only 
present in S/S{'} systems, which comes from a difference in
diffusion constants for the two superconductors, and
to which we do not pay further attention.\cite{aarts0,aarts}

Up to now, the occurrence of the dimensional crossover has not been described
satisfactorily. The calculated crossover temperature appears
to lie much higher than the measured one.\cite{rutg} But
the results obtained so far only apply for infinite periodic multilayers.
Such model systems have translational symmetry in the growth
direction, so that surface effects are excluded. It was Aarts who
suggested to investigate finite size effects, which always may
show up in the finite samples used in experiments.\cite{aarts0}

An extension of the applications of the theory to finite systems
is not a trivial one. The reason is, that in calculating 
the parallel upper critical magnetic field one has to consider
the position of the nucleation point $z_0$.
In infinite periodic systems, 
nucleation always occurs in the middle of a layer, due to 
the periodical symmetry of the system. 
Each center of a layer is a mirror-symmetry center.
One only has to figure out which layer of the 
two types of layers will correspond to the optimal nucleation.
In practice this has led to a distinction of two possible
solutions, a first solution corresponding to nucleation
in the S layers, and a second solution describing nucleation 
in the N layers.\cite{rutg}

For finite multilayers the symmetry is largely reduced. Due
to mirror-symmetry with respect to the center of the system,
one would be inclined to implement the choice of 
the nucleation point in the center only, as we
did in the previous section. But finite
systems are known to exhibit surface superconductivity
as well. This means that one has to search for the precise
location of the nucleation point. If it would turn out
to lie close to the surface, it would lie at
a point devoid of any symmetry. While surface superconductivity
is well investigated in finite homogeneous superconductors,
as far as the authors know, such a search has never been done 
for inhomogeneous layered systems. This is the subject of the present
section.

In looking for the correct solution of Eq. (\ref{lambda}) it
is important to make use of the fact, that the eigenfunctions (\ref{psilampar})
corresponding to different nucleation points are
orthogonal, since they have different wave-numbers $k_x$.
Due to this, 
they provide zero-matrix elements in the matrix $V_{\lambda \lambda '}$. 
The determinant (\ref{det}) will be split in an infinite number of blocks, 
each block corresponding to a certain nucleation point $z_0=\xi^2 k_x$. 

\begin{equation}
\label{det-z0}
\det|\delta _{\lambda \lambda '}-2\pi kT\sum _{\omega }\frac{1}{2|\omega |+
\epsilon_{\lambda }}V_{\lambda \lambda '}|=\nonumber\\
\prod_{z_0}{\det|\delta _{\lambda,z_0 ;\lambda ',z_0}-2\pi kT\sum _{\omega }\frac{1}{2|\omega |+
\epsilon_{\lambda,z_0 }}V_{\lambda,z_0 ;\lambda ',z_0}|}=
0.
\end{equation}
In the determinant equation (\ref{det-z0}), we are interested to solve for 
that block 
\begin{equation}
\label{det-param}
\det|\delta _{\lambda,z_0 ;\lambda ',z_0}-2\pi kT\sum _{\omega }\frac{1}
{2|\omega |+\epsilon_{\lambda,z_0 }}V_{\lambda,z_0 ;\lambda ',z_0}|=0,
\end{equation}
which has the highest $T_c$ as solution for  the critical temperature. 
This will be the physical solution.

\begin{figure}[htb]
\centerline{\epsfig{figure=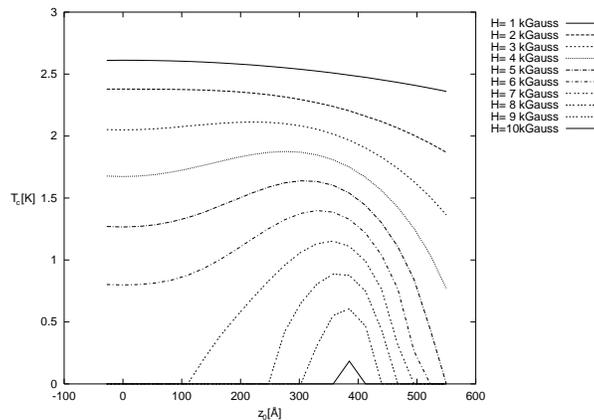,angle=-90,width=8.0cm}}
\caption[]{The critical temperature versus the nucleation point, measured from
the center of the layer. 
The magnetic field varies from 1 to 10 kGauss. Due to the 
symmetry of the multilayer, we only show one half of the system.}
\label{nucl-surf}
\end{figure}

Results for an 11-layers system, obtained according to this scheme
are shown in FIG. \ref{nucl-surf}. The critical
temperature $T_c$ is plotted as
a function of the position of the nucleation point.
The system parameters used apply for a V(100\AA)/Ag(100\AA) system,
having a $H_{c2,\parallel}(0)$ of about 11 kGauss.
The specific set of parameteres is given in the literature. 
\cite{kanoda1,kanoda2,kanoda3,kanoda4}

For a small magnetic field, for 1 and 2 kGauss, the superconductivity occurs 
in the middle of the multilayer, as the highest $T_c$ is found for
the nucleation point right in the center of the system. For such a small
field the size of the system is of the same order of magnitude as
$\xi=\sqrt{\frac{\hbar c}{2eH}}$, which
quantity gives the scaling of the order parameter. For
1 kGauss $\xi = 574 $ \AA, which is relatively large
compared to the layer thickness.
Microscopically, at low parallel 
magnetic field the pair amplitude is spread over the 
entire system and looks the same for any choice of the nucleation
point $z_0$. One could say as well, that for such small fields the entire
system is just a surface. At higher parallel 
magnetic fields, the pairs are confined in about one layer around 
the position of $z_0$. This makes that for higher
fields a real difference can
be expected for two different $z_0$ choices. For $H \geq$ 3 kGauss,
the nucleation point shifts from the center of the system towards 
the surface of the sample. For high fields one clearly sees,
that superconductivity is found at the surface of the system only.

\begin{figure}[htb]
\centerline{\epsfig{figure=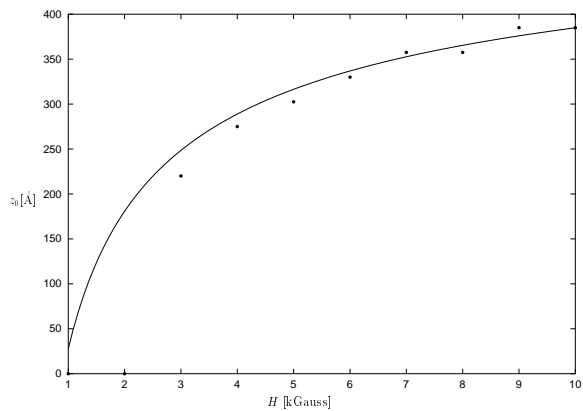,angle=90,width=8.0cm}}
\caption[]{
The points in which the nucleation occurs for different magnetic fields 
(with dots) 
fitted in terms of the magnetic coherence length 
$\xi$ (continuous line). 
}
\label{surface}
\end{figure}

The shift of the nucleation point towards the surface with
increasing magnetic field illustrates the scale of the surface nucleation,
which is given by the magnetic coherence length.
For a bulk superconductor, if the magnetic field is applied parallel to the 
interface with the vacuum, the pair function is localized around $z_0$ 
nearby the surface\cite{degennes}
\begin{equation}
F\cong exp[-\frac{1}{2}(\frac{z-z_0}{\xi })^2],
\label{dege}
\end{equation}
in which $z$ is now measured from the surface, situated at $z=0$.
The expression (\ref{dege}) is a gaussian trial function which shows 
that the pair function is localized around the nucleation point $z_0$. 
A detailed calculation which is based on the exact Weber-function solution 
shows that $z_0\sim\xi$ and the optimal value is $z_0 = 0.59~\xi$. 

For our multilayer system, the dots in FIG. \ref{surface} show 
the nucleation point as a function of the magnetic field, which
can be converted in a dependence on the magnetic coherence length.
Measuring $z_0$ from the surface we find $z_0 = C \xi$,
where $C=0.91$, which is represented in FIG. \ref{surface}
by the continuous line. It is not too surprising that this
constant for the layered system is larger than the value
of 0.59 for a bulk superconductor. The surface phenomena in the present
systems will be shifted inward due to the covering $N$ layer.

\begin{figure}[htb]
\centerline{\epsfig{figure=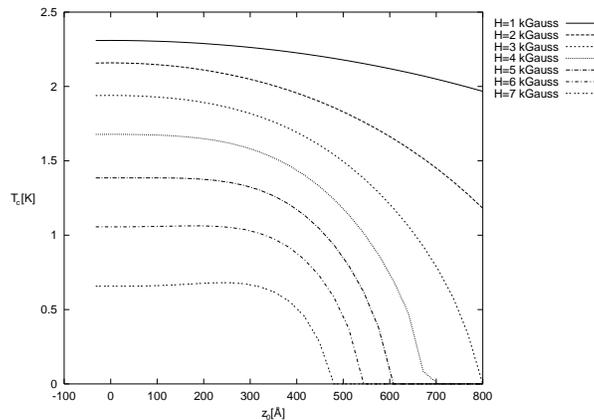,angle=-90,width=8.0cm}}
\caption[]{For a thick exterior N layer, we have bulk superconductivity,
rather than surface superconductivity. }
\label{thickN}
\end{figure}

The results presented up to now in this section
apply to a multilayer system whose outer N layers have a thickness 
of 100\AA, which is equal to the thickness of the S and N internal layers. 
As we were looking at the influence of the boundary on the superconductivity,
we studied also systems in which the thickness of the outer N layers is larger.
Keeping all other parameters constant, we only extend the thickness 
of the exterior layers to $d_N^{\rm outer}=350$ \AA, such that it 
becomes larger than the magnetic coherence length $\xi$.
The corresponding plot of $T_c$ versus $z_0$ is given in
FIG. \ref{thickN} for seven magnitudes of the magnetic field.
Clearly, up to 5 kGauss the nucleation occurs in the
middle of the system. No surface superconductivity does
appear anymore since the normal outer layer
is thick enough to ``screen'' it. For larger fields some
surface superconductivity shows up again.

In conclusion, in spite of the complexity of the
layered systems, relations found
between the nucleation and the scaling length $\xi$ of the superconductivity
compare well with established results for homogeneous superconductors.
This can be regarded as a support for the validity
of the theory of Takahashi and Tachiki, which is
devised to describe arbitrary inhomogeneous systems.

\section{The influence of the parameters on the proximity effect}
\label{parameters}

Up to now we have shown properties of finite multilayer
systems by varying the applied magnetic field only. In order
to prepare a solution to the as yet unsolved problem
related to the position of the
dimensional crossover in the $H_{c2,\parallel}(T)$ curves,
we need to study the way in which nucleation occurs
by varying system parameters as well. In addition to
the system parameters mentioned already, such as
the diffusion constant for S and N layers and the density of states
(DOS) of the S layer, we will consider the resistivity of the interfaces.
\cite{kuprianov,rutger2}
Although we start from parameters which are specific to V/Ag,
by varying some of them the set 
we are using might not characterise this system anymore. On the other hand, 
we are studying general properties of S/N systems, which means that we 
have to look not only to a particular system, but to a large class 
of S/N multilayers. 

\begin{figure}[htb]
\centerline{\epsfig{figure=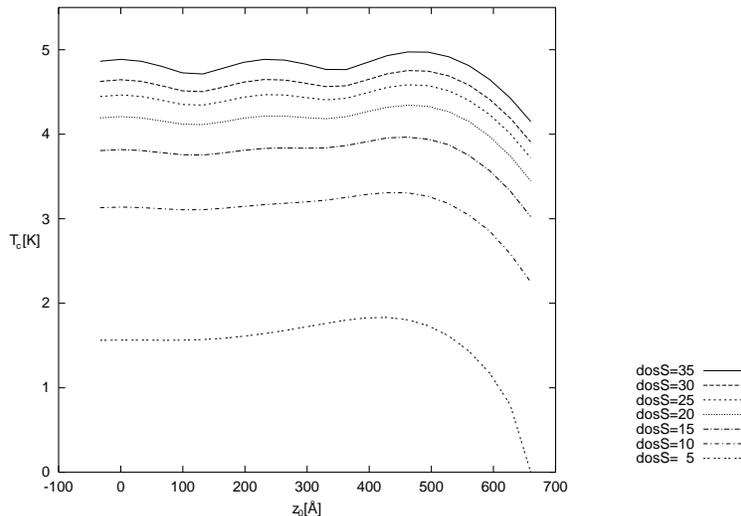,angle=-90,width=10.0cm}}
\caption[]{The nucleation curves for a 11-layers S(120\AA)/N(120\AA) 
system, with increasing the S layer DOS. 
H=6 kGauss and $D_N=5$ cm$^2$/s.}
\label{dosS}
\end{figure}

\begin{figure}[htb]
\centerline{\epsfig{figure=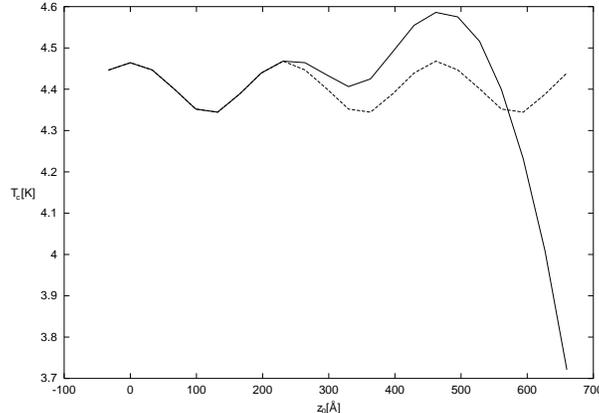,angle=-90,width=8.0cm}}
\caption[]{Nucleation curves for systems which have the same set 
of parameters. The dashed line corresponds to the infinite periodical 
multilayer, while the continuous one to a finite 11 layer system. }
\label{infinite}
\end{figure}

First we vary the S layer DOS in
a S(120\AA)/N(120\AA) 11-layers system. This is a parameter which 
measures the density of electrons at the Fermi level. The higher it is, 
the more Cooper pairs can be formed, so the material is a better 
superconductor. While varying the S layer DOS, the other parameters
keep the same values used before.
In FIG. \ref{dosS} we show the nucleation curves 
for an increasing DOS of the superconducting layer.
The quantity dosS gives the ratio of the S layer DOS
with respect to the N layer DOS, which covers a large range of
DOS values, because in the equations only this ratio contributes.
Since the system is symmetric with respect to the center of the multilayer,
we only show half of it. 
First, we notice an increase of the multilayer critical temperature 
with increase of the S layer DOS, which is a consistent result.
Secondly, with the increase of the S layer DOS, 
we notice a smooth transition from a regime which is dominated by 
the surface superconductivity, 
at lower dosS values, to a regime in which a modulation develops
following the structure of the multilayer.
Besides, surface superconductivity is still present by a slight bump close
to the surface. 
This bump cannot be present in infinite multilayers, having no surface. 
This is illustrated in FIG. \ref{infinite}, in which the
dosS = 25 result for the
finite system, represented by the solid line, is compared to
the periodic result for an infinite multilayer.
The modulations are such, that the maxima are located
at the centers of the superconducting layers. Apparently,
for high dosS values the proximity effect is not strong enough
anymore to smear out the high pair amplitude inside the S layers. 
This can be understood from the fact that at high dosS, 
the S/N interfaces behave like S/Vacuum interfaces, so that the coupling 
between the S-layers is very small. 

\begin{figure}[htb]
\centering
\begin{tabular}{c}
\epsfig{figure=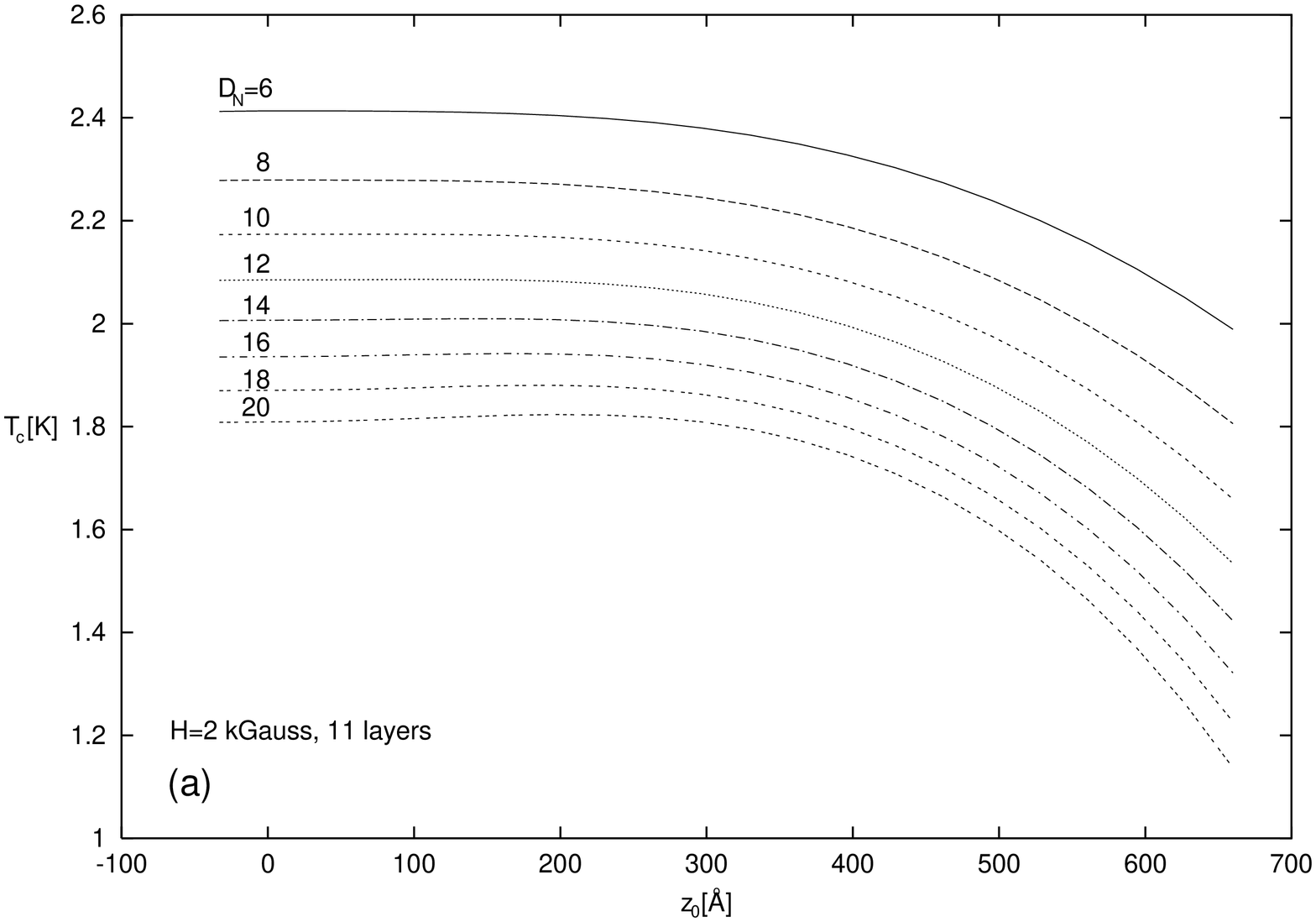,angle=-90,width=8.0cm}
\\
\epsfig{figure=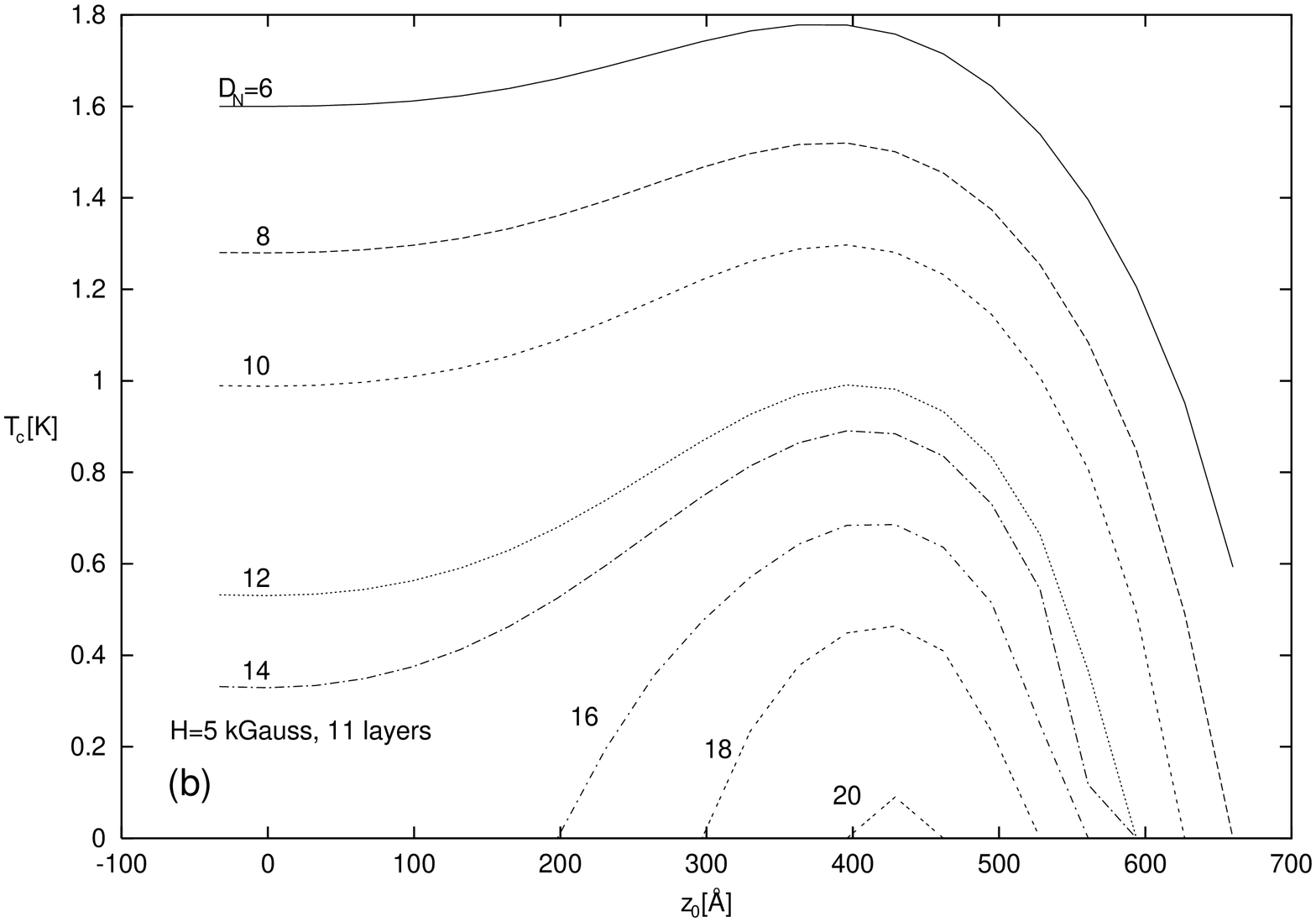,angle=-90,width=8.0cm}
\\
\epsfig{figure=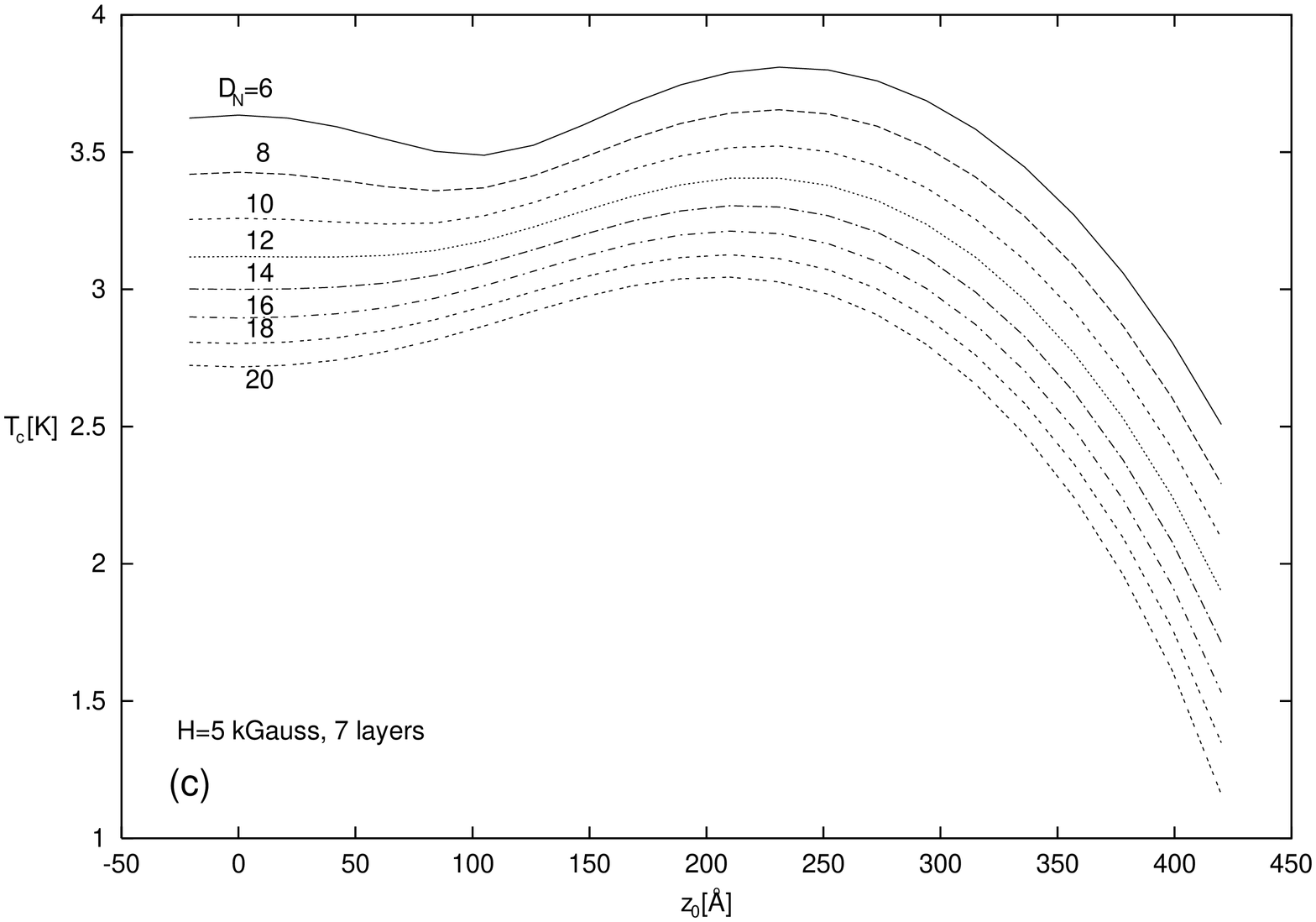,angle=-90,width=8.0cm}
\end{tabular}
\caption{The temperature as a function of the nucleation point $z_0$,
with increasing diffusion coefficient of the normal metal $D_N$.
(a)at H=2 kGauss, $\frac{\rm S layer DOS}{\rm N layer DOS}=5$; 
(b)H=5 kGauss, $\frac{\rm S layer DOS}{\rm N layer DOS}=5$; 
(c)H=5 kGauss, $\frac{\rm S layer DOS}{\rm N layer DOS}=25$.}
\label{difN_grup}
\end{figure}

Now we want to see how the behaviour displayed in FIG. \ref{dosS}
is modified by varying the diffusion constant $D_N$ of the N layer
and by looking at different magnitudes of the magnetic field.
We concentrate on two representative curves, namely for dosS =
5 and 25. Results are shown in FIG. \ref{difN_grup}.

In FIG. \ref{difN_grup}(a) we show the evolution of the 
nucleation curve for dosS = 5 with increasing
diffusion coefficient $D_N$,
at a lower magnetic field (H=2 kGauss). Even the bump in the
dosS = 5 curve of FIG. \ref{dosS} has disappeared, because
for such a weak field $\xi$ is large, and
there is no surface superconductivity.
For larger $D_N$ values some surface superconductivity shows
up again. This is because, in addition to the magnetic coherence length,
from now on to be denoted by $\xi_{mag}$,
the BCS coherence length comes into play. For dirty alloys
this quantity is defined as 
$\xi_{BCS}=\sqrt{\frac{\hbar D}{2\pi kT}}$,\cite {degen}
and for D$_N$ = 20~{cm}$^2$/s and T = 2~K it is equal to 348 \AA.
This implies a good coupling between the superconducting layers, by
which the entire multilayer develops properties pointing in the
direction of a homogeneous superconductor. Because
$\xi_{mag}=406$ \AA, for such a system surface superconductivity
could develop at $0.59\times406 = 239$ \AA ~from the surface.
The weak maximum in the $D_N = 20$ curve in FIG. \ref{difN_grup}(a)
lies somewhat more inward.

In FIG. \ref{difN_grup}(b) the magnetic field has a higher value, 
H=5 kGauss, which lies close to the value used in FIG. \ref{dosS}.
Again dosS = 5, and the curve for D$_N$ = 6 is similar to
the dosS = 5 curve of FIG. \ref{dosS}. The effect of
increasing $D_N$ is much more pronounced than for the
weaker field,  and the maxima of the curves lie at a distance
of about $0.91 \xi_{mag}=234~{\rm \AA}$ from the surface. 
For large $D_N$ values
surface superconductivity is dominantly present.

FIG. \ref{difN_grup}(c) shows the modifications to the
dosS = 25 curve of FIG. \ref{dosS}. We restricted the calculation
to a 7-layer system, by which one oscillation has dropped out.
The magnetic field is H = 5 kGauss, slightly weaker than the
field used in FIG. \ref{dosS}.
The modulation disappears with increasing diffusion 
coefficient, because a higher $D_N$ facilitates the coupling between 
the S-layers. 

All three panels of FIG. \ref{difN_grup} show a decrease of
$T_c$ with an increasing diffusion coefficient.
This can be understood globally in terms of superconducting
bulk properties. For a bulk superconductor the only eigenvalue
that contributes in the determinant (\ref{det}) is the
ground state eigenvalue, which is proportional to the diffusion
constant and the magnetic field, and inversely proportional to
the magnetic flux quantum $\phi_0 = \frac{hc}{2e}$,\cite{taka,degennes,lodkop}
\begin{equation}
\label{eGDH}
\epsilon _G=\frac{2\pi\hbar DH}{\phi_0}.
\end{equation}
Following de Gennes we write $\epsilon _G$ in Eq. (\ref{det})
explicitly in terms of the so-called universal function {\rm Un(x)} 
as\cite{degennes}
\begin{equation} 
\label{eGdeGennes}
\epsilon _G=kT_c(0){\rm Un}\left(\frac{T_c(H)}{T_c(0)}\right),
\end{equation}
while {\rm Un(x)} is monotonically increasing with decreasing $x$.
This implies a decrease
of $T_c(H)$ with increasing $\epsilon _G$,
and therefore, according to Eq. (\ref{eGDH}), with increasing $D$. Also,
for a larger magnetic field the effect will be stonger. 
For the layered systems an average value of $D_S$ and $D_N$
has to be used, but this average increases with increasing
$D_N$. A decrease of $T_c$ with increasing $D_N$ is seen indeed,
and also the effect is stronger for a larger field.

\begin{figure}[htb]
\centerline{\epsfig{figure=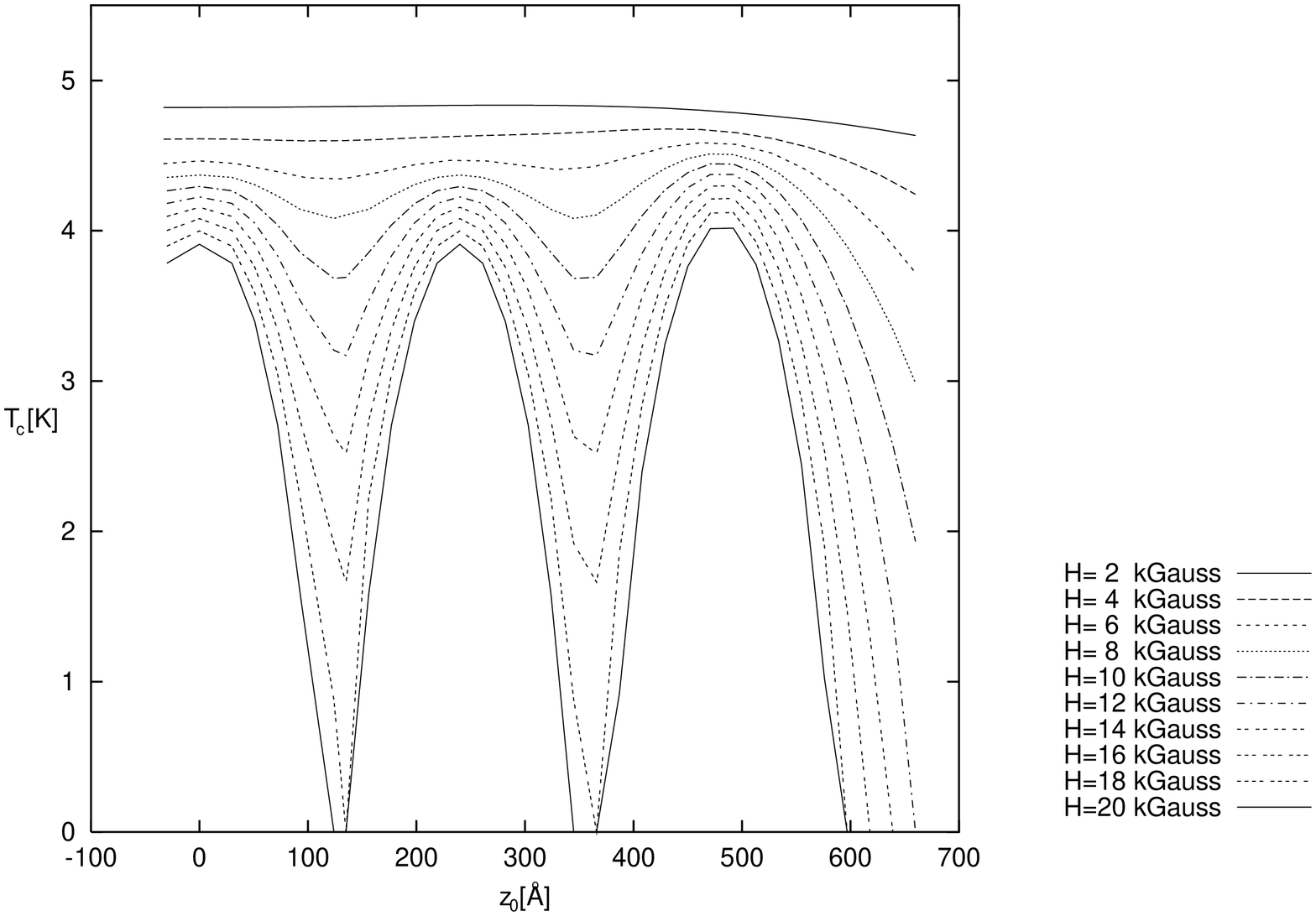,angle=-90,width=10.0cm}}
\caption[]{The temperature curves for different nucleations,
with increasing the external applied magnetic field H. }
\label{dosS25_H}
\end{figure}

The scaling change due to the variation of $\xi_{mag}$ through 
its dependence on $H$ is shown in FIG. \ref{dosS25_H}. The
$H = 6$ kGauss curve reproduces the dosS = 25 curve of FIG. \ref{dosS}.
The nucleation curves show
a smooth transition from curves which exhibit only surface 
superconductivity, to curves in which a modulation over the 
S and N layers is present. 
This transition happens when H$\approx$ 5 kGauss and $\xi _{mag}\approx
257$ \AA. This value is comparable with the periodicity 
of the multilayer, $d_{\rm N}+d_{\rm S} = 240$ \AA. 
In other words, when $\xi _{mag}$ becomes as small as the periodicity 
of the multilayer, the nucleation exhibits the modulation corresponding 
to the geometrical structure of the multilayer. 
The maxima are located in the S layers and a preference is
shown for the superconductivity to nucleate there. 
Comparing with Fig. \ref{dosS}, in which this modulation appears 
when the coupling between the S-layers becomes weak at high dosS, 
here this feature of the nucleation curves can again be understood 
by the fact that the S-layer coupling decreases with the decrease of the 
coherence length at high magnetic fields. 

\begin{figure}[htb]
\centerline{\epsfig{figure=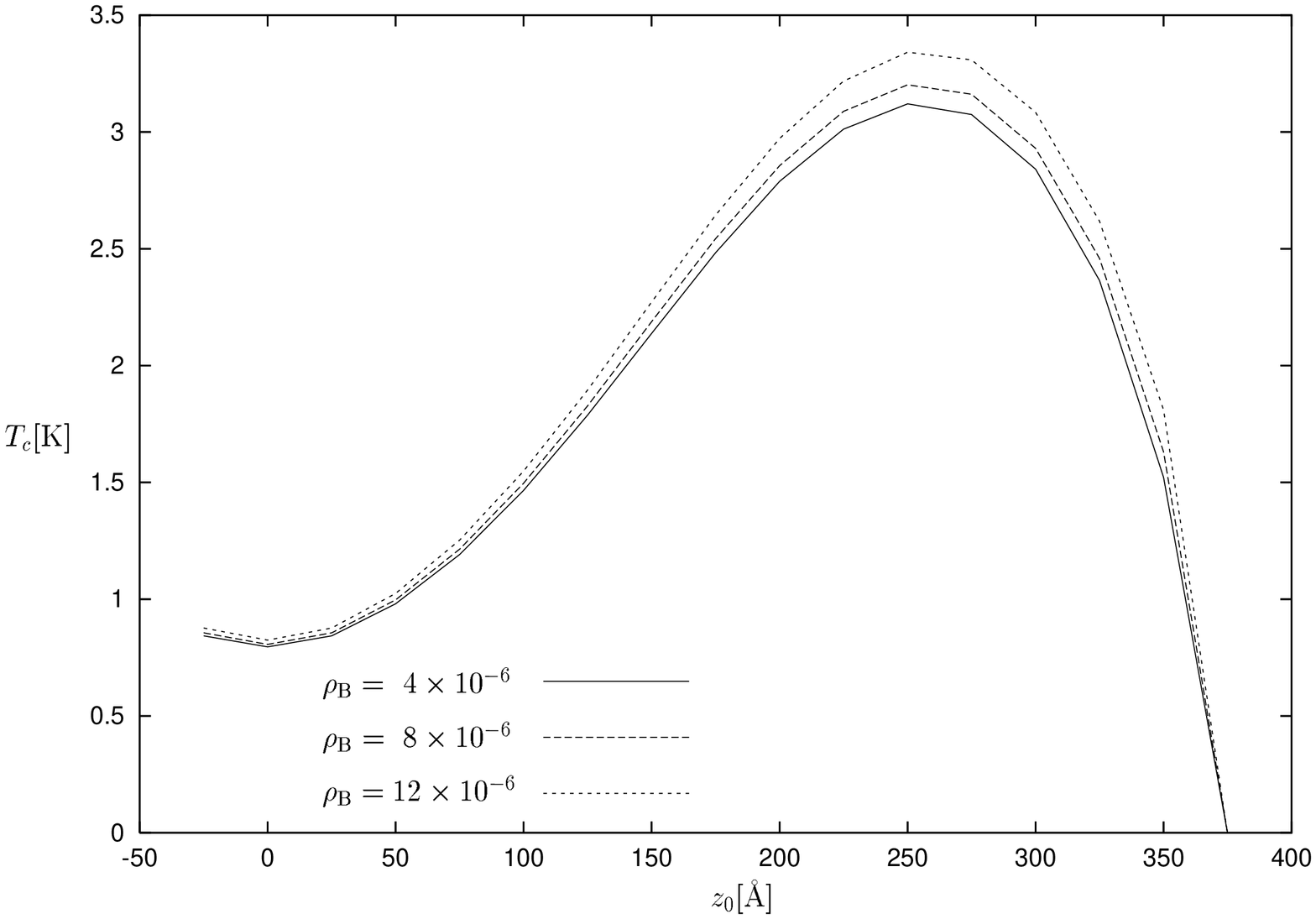,angle=90,width=10.0cm}}
\caption[]{The nucleation curves for different nucleations,
with increasing boundary resistivity ${\rho}_{\rm B}$($\mu \Omega {\rm cm}^2$).
The system is a thin film of Nb covered with Pd. The magnetic field 
applied is H=12.5 kGauss.}
\label{transp}
\end{figure}

Finally we vary the boundary resistivity ${\rho}_B$ at the 
interfaces, expressed in units $\mu\Omega{\rm cm}^2$. 
Since we have in view interfaces, 
we restrict our calculations to just 
a thin superconductor film, covered by a normal metal. 
We vary ${\rho}_B$ around experimental data, \cite{kaneko} which 
are specific to Nb/Pd systems. We did calculations 
for a thin film of Nb(600\AA) covered by Pd(100\AA) 
at an intermediate magnetic field, which for Nb/Pd is about 12.5 kGauss. 
The nucleation curves are shown in FIG. \ref{transp}. 
We notice that with increasing boundary resistivity ${\rho}_B$,
which means that the S/N interfaces become less transparent to the
electrons, the critical temperature increases. This can be understood
as follows. The interface resistivity
makes the spreading of the electrons (Cooper pairs)
over the different layers more difficult, which effectively
leads to an enhancement of the pair amplitude in the superconducting
layers. This weaker proximity effect results in a higher $T_c$
of the system. This effect is similar to what was discussed
in Sec. \ref{theory} regarding $H_{c2,\parallel}$, which systematically
turned out to be larger than $H_{c2,\perp}$.

\section{Conclusions} 
\label{concl}

The theory of Takahashi and Tachiki, applied so far either to a
bilayer\cite{koperN} or to infinite multilayers,\cite{taka,rutg}
has been applied to describe superconducting properties of
finite multilayers. We have calculated the spatial dependent
pair amplitude for two magnetic field directions, parallel
and perpendicular to the layers. These properties, including the
upper critical field curves, have been studied as a function of the
number of layers.

The nucleation problem, which appears when one studies 
systems to which a parallel magnetic field is applied, is illustrated on 
model systems such as V/Ag multilayers and Nb/Pd thin films for
a large range of parameters. Moreover, we compare the results obtained 
for finite systems to what is found for infinite periodic multilayers.
Since we take into 
account the boundaries, typical for a finite sample, we get surface
superconductivity. 

Varying  the parameters we see how they influence the proximity 
effect and the scaling of the sample and layer thicknesses
to the magnetic and BCS coherence lengths.
The superconductivity occurs differently for different choices 
of the parameters. 
If we define the dimensional crossover as being the temperature 
where the modulation over the geometrical structure of the multilayer 
appears in the 
nucleation curves, then 
we must expect a shift of the dimensional crossover in the  
$H_{c2,\parallel}(T_c)$ curves 
under a change of the parameters. 

The results we get are meant to prepare an improvement upon the previous approaches 
in explaining the experimental results.\cite{rutg} 
In addition, the results obtained contribute to new confidence in the
Takahashi-Tachiki theory. This comes from the observation,
that our results about surface superconductivity in finite multilayers,
being typically anisotropic systems,
can be understood as a generalization of what is known about the
phenomenon in isotropic homogeneous superconductors.

\section*{acknowledgement}

One of the authors (CC) would like to thank Dr. R.T.W. Kooperdraad for 
useful discussions.

\end{document}